\documentclass[english]{article}
\usepackage[T1]{fontenc}
\usepackage[latin9]{inputenc}
\usepackage{array}
\usepackage{booktabs}
\usepackage{multirow}
\usepackage{amsmath}
\usepackage{amsthm}
\usepackage{amssymb}

\makeatletter

\providecommand{\tabularnewline}{\\}

\numberwithin{equation}{section}
\numberwithin{figure}{section}

\usepackage{babel}
\usepackage{fullpage}
\usepackage{authblk}
\usepackage{hyperref}
\usepackage{placeins}
\usepackage{hyphenat}
\usepackage{fancyvrb}
\usepackage[hypcap]{caption}

\usepackage[title,titletoc]{appendix}

\title{Computing Credit Valuation Adjustment solving coupled PIDEs in the Bates model}
\author{ \textsc{Ludovic Gouden\`ege}\thanks{F\'ederation de Math\'ematiques de CentraleSup\'elec - CNRS FR3487, France -\texttt{ ludovic.goudenege@math.cnrs.fr}} \and \textsc{Andrea Molent}\thanks{Universit\`a degli Studi di Udine, Italy - \texttt{andrea.molent@uniud.it}} \and \textsc{Antonino Zanette}\thanks{Dipartimento di Scienze Economiche e Statistiche, Universit\`a degli Studi di Udine, Italy - \texttt{antonino.zanette@uniud.it}}}
\date{}

\makeatother

\usepackage{babel}
\begin{document}
\maketitle

\begin{flushleft}
\rule{1\columnwidth}{1pt}
\par\end{flushleft}

\begin{flushleft}
\textbf{\large{}Abstract}
\par\end{flushleft}{\large \par}

Credit value adjustment (CVA) is the charge applied by financial institutions
to the counterparty to cover the risk of losses on a counterpart default
event. In this paper we estimate such a premium under the Bates stochastic
model (Bates \cite{bates1996}), which considers an underlying affected
by both stochastic volatility and random jumps. We propose an efficient
method which improves the finite-difference Monte Carlo (FDMC) approach
introduced by de Graaf et al. \cite{deGraaf2017}. In particular,
the method we propose consists in replacing the Monte Carlo step of
the FDMC approach with a finite difference step and the whole method
relies on the efficient solution of two coupled partial integro-differential
equations (PIDE) which is done by employing the Hybrid Tree-Finite
Difference method developed by Briani et al. \cite{briani2016,briani2017,briani2015}.
Moreover, the direct application of the hybrid techniques in the original
FDMC approach is also considered for comparison purposes. Several
numerical tests prove the effectiveness and the reliability of the
proposed approach when both European and American options are considered. 

\vspace{2mm}

\noindent \emph{\large{}Keywords}: Credit Value Adjustment, Hybrid
methods, PIDE, Monte Carlo, Bates model.

\noindent\rule{1\columnwidth}{1pt}

\newpage

\section{Introduction}

Financial institutions suffer several risks, one of them is the counterparty
credit risk (CCR). This risk arises from the possibility that the
counterparty of a financial contract may default. This risk was often
overlooked, but in the last decades, after the financial crisis of
2007 and the Lehman Brothers failure in 2008, it gained more and more
interest by practitioners and academics. In particular, according
to the Basel III framework of 2010, financial institutions must charge
a premium to their counterparty according to its credit reliability
in order to compensate for a possible counterparty default. Also IFRS
13 in 2013 requires the fair value of financial products to be measured
based on counterparty credit risk. For these reasons, financial institutions
charge to the counterparty a premium called Credit Valuation Adjustment
(CVA), which is the difference between the risky value and the current
risk-free value of derivatives contract. Estimating the good value
of the CVA can be a demanding effort due to its complicated definition
and to its dependence by the underlying stochastic model. As no specific
method is prescribed in the accounting literature, various approaches
are used in practice by derivatives dealers and by end users to estimate
the effect of credit risk on the fair value of financial derivatives.
The common approach is to price the CVA through the so called expected
exposure, which is the mean exposure distribution at a future date.
Usually, this exposure is calculated by means of Monte Carlo approaches
which are computationally expensive. In particular, nested Monte Carlo
simulations or least squares techniques are employed by Joshi and
Kwon \cite{joshi2016}. These techniques have been improved through
the use of stochastic grid bundling method by Jain and Oosterlee \cite{jain2015}
and by Karlson et al. \cite{karlsson2016}. An interesting approach
to compute the CVA - when the Heston model is assumed - is the so
called finite-difference Monte Carlo (FDMC) method, proposed by de
Graaf et al. \cite{deGraaf2017}. Such an approach combines the finite-difference
method and the Monte Carlo method to solve a partial differential
equation (PDE) and to estimate the mean exposure respectively. Feng
\cite{feng2017} adapted the FDMC approach to deal with the case of
an underlying evolving according to the Bates model: in this particular
case, the PDE to be solved is replaced by a partial integral differential
equation (PIDE), which implies an additional computational effort.
Other recently introduced alternative approaches consist in employing
the fast Fourier transform (Borovykh et al. \cite{borovykh2018})
or marked branching diffusions (Henry-Labord\`ere \cite{labordere2012}). 

In this paper we focus on the computation of the CVA when the Bates
model is considered and we propose an efficient method which improves
the results of the FDMC method. Specifically, the Bates model considers
an underlying affected by both stochastic volatility and random jumps:
the dynamics of the underlying asset price is driven by both a Heston
stochastic volatility \cite{heston1993} and a compound Poisson jump
process of the type originally introduced by Merton \cite{merton1976}.
Such a model was introduced by Bates \cite{bates1996} in the foreign
exchange option market in order to tackle the well-known phenomenon
of the volatility smile behavior. In the case of plain vanilla European
options, Fourier inversion methods, employed by Carr and Madan \cite{carr1999},
lead to closed-form formulas to compute the price under the Bates
model. Two innovative and efficient approaches to price derivatives
when the Bates model is considered are proposed by Briani et al. \cite{briani2016},
the so called Hybrid Tree-Finite Difference method and Hybrid MC method. 

Our main result consists in developing an efficient numerical method
to estimate the CVA. In particular, the method we propose consists
in computing the CVA value as the solution of two coupled PIDE \textendash{}
one for the risk free price and one for the risk adjusted price \textendash{}
which are solved by means of the Hybrid Tree-Finite Difference method.
That is, we replace the Monte Carlo step in the FDMC method with the
computation of the solution of a PIDE and this improves clearly the
computational efficiency. 

Moreover, the direct application of the hybrid techniques in the original
FDMC approach is considered for comparison purposes. Specifically,
we apply the two hybrid methods developed by Briani et al. to perform
both the finite difference step and the Monte Carlo step of the FDMC
approach. 

Several numerical experiments show that the proposed method is efficient
and reliable as it provides accurate approximations for the CVA value
with a low computational cost. 

The reminder of the paper is organized as follows. Section 2 introduces
the Bates stochastic model and the partial integro-differential equation
that allows one to compute option prices. Section 3 presents the CVA
definition and some useful properties used in the following. Section
4 outlines the main features of the hybrid methods and how to employ
them in option pricing. Section 5 describes the numerical methods
for computing the CVA. Section 6 presents and discusses the results
of the numerical simulations. Section 7 concludes.

\section{The Model}

In the Bates model the volatility is assumed to follow the CIR process
(Cox et al. \cite{cox2005}) while the underlying asset price process
contains a further noise from a jump process. In particular, the model
for the stock price and its volatility is given by the following relations:
\begin{equation}
\begin{array}{l}
{\displaystyle \frac{dS_{t}}{S_{t^{-}}}=(r-\eta)dt+\sqrt{V_{t}}\,dZ_{t}^{S}+dH_{t},\smallskip}\\
dV_{t}=\kappa(\theta-V_{t})dt+\sigma\sqrt{V_{t}}\,dZ_{t}^{V},\smallskip
\end{array}\label{BHHmodel}
\end{equation}
 where $\eta$ denotes the continuous dividend rate, $S_{0}$, $V_{0}$
are positive values, $Z^{S}$, $Z^{V}$ are correlated Brownian motions
such that $\left\langle dZ_{t}^{s},dZ_{t}^{V}\right\rangle =\rho dt$
and $H$ is a compound Poisson process with intensity $\lambda$ and
i.i.d. jumps $\{J_{k}\}_{k}$, that is 
\begin{equation}
H_{t}=\sum_{k=1}^{K_{t}}J_{k},\label{H}
\end{equation}
$K$ denoting a Poisson process with intensity  $\lambda$. As in
the model of Merton \cite{merton1976}, we assume that the jumps are
i.i.d. log-normal random variables, described by 
\begin{equation}
\log\left(1+J\right)\sim N\left(\alpha-\frac{\beta^{2}}{2},\beta^{2}\right),\label{eq:log_jumps}
\end{equation}
with $\alpha$ and $\beta$ real parameters.

We consider an option with maturity $T>0$, payoff function $\psi\left(S\right)$
and we denote with $\mathcal{V}\left(t,S,V\right)$ its fair price
at time $t$, if $S_{t}=S$ and $V_{t}=V$. Then, the price of the
derivative $\mathcal{V}$ is then given by 
\begin{equation}
\mathcal{V}\left(t,S,V\right)=\mathbb{E}\left[e^{-r\left(T-t\right)}\psi\left(S_{T}\right)|S_{0}=S,V_{0}=V\right].\label{eq:risk-free-price}
\end{equation}

Using a martingale approach for an European option, it is possible
to show that $\mathcal{V}\left(t,S,V\right)$ satisfies the following
partial integro-differential Equation (PIDE) (Salmi et al \cite{salmi2014}):
\begin{multline}
\frac{\partial\mathcal{V}\left(t,S,V\right)}{\partial t}+\frac{1}{2}VS^{2}\frac{\partial^{2}\mathcal{V}\left(t,S,V\right)}{\partial S^{2}}+\rho\sigma VS\frac{\partial^{2}\mathcal{V}\left(t,S,V\right)}{\partial S\partial V}+\frac{1}{2}\sigma^{2}V\frac{\partial^{2}\mathcal{V}\left(t,S,V\right)}{\partial V^{2}}+\left(r-q-\lambda\left(e^{\gamma}-1\right)\right)\frac{\partial\mathcal{V}\left(t,S,V\right)}{\partial S}\\
+\kappa\left(\theta-V\right)\frac{\partial\mathcal{V}\left(t,S,V\right)}{\partial V}-\left(r+\lambda\right)\mathcal{V}\left(t,S,V\right)+\lambda\int_{0}^{\infty}\mathcal{V}\left(t,xS,V\right)p_{J}\left(x\right)dx=0,\label{eq:PIDE}
\end{multline}
with the terminal condition is 
\begin{equation}
\mathcal{V}\left(T,S,V\right)=\psi\left(S\right),
\end{equation}
where $p_{J}\left(x\right)$ is the log-normal probability density
function of the jump variable $J$. 

We define the following operators: 

\begin{align}
L_{C}\mathcal{V}= & \frac{1}{2}VS^{2}\frac{\partial^{2}\mathcal{V}}{\partial S^{2}}+\rho\sigma VS\frac{\partial^{2}\mathcal{V}}{\partial S\partial V}+\frac{1}{2}\sigma^{2}V\frac{\partial^{2}\mathcal{V}}{\partial V^{2}}+\left(r-q-\lambda\left(e^{\gamma}-1\right)\right)\frac{\partial\mathcal{V}}{\partial S}+\kappa\left(\theta-V\right)\frac{\partial\mathcal{V}}{\partial V}-\left(r+\lambda\right)\mathcal{V},\label{eq:LCV}
\end{align}
and
\begin{equation}
L_{J}\mathcal{V}=\lambda\int_{0}^{\infty}\mathcal{V}\left(t,xS,V\right)p_{J}\left(x\right)dx.\label{eq:LJV}
\end{equation}
Therefore, by using (\ref{eq:LCV}) and (\ref{eq:LJV}), the PIDE
(\ref{eq:PIDE}) can be rewritten in the more compact form as follows:
\begin{equation}
\frac{\partial\mathcal{V}\left(t,S,V\right)}{\partial t}+L_{C}\mathcal{V}\left(t,S,V\right)+L_{J}\mathcal{V}\left(t,S,V\right)=0.\label{eq:PIDE_compact}
\end{equation}

\section{The CVA}

Let us define the exposure (or financial exposure) $E\left(t\right)$
towards a counterparty at time $t$ as the positive side of a contract
(or portfolio) value $\mathcal{V}\left(t,S_{t},V_{t}\right)$, that
is 
\begin{equation}
E\left(t\right)=\max\left[\mathcal{V}\left(t,S_{t},V_{t}\right),0\right].
\end{equation}
This amount represents the maximum loss if the counterparty defaults
at $t$: an economic loss would occur if the transactions with the
counterparty has a positive economic value at the time of default
(see the Basel Committee \cite{comitato2004}).

Let us define the present Expected Exposure (EE) at a future time
$t<T$ as 
\begin{equation}
EE\left(t\right)=\mathbb{E}\left[E\left(t\right)|\mathcal{F}_{0}\right]\label{eq:EE}
\end{equation}
where $\mathcal{F}_{0}$ is the filtration at time $t=0$. In particular,
in the case of a long position in a Call or Put option, the price
is always positive and thus the EE is equal to the future option price.

Let $\tau_{C}$ denote the counterparty's default time: such a time
is a random variable which is supposed to be independent from the
stochastic processes $Z^{S}$, $Z^{V}$ and $H$. Moreover, we describe
its cumulative distribution function as follows:
\begin{equation}
PD\left(t\right)=1-\exp\left(-\int_{0}^{t}\delta\left(s\right)ds\right).
\end{equation}
Here, $\delta\left(t\right)$, the so called hazard rate, is a non-negative
function such that $\int_{0}^{+\infty}\delta\left(s\right)ds=+\infty$
(Promislow \cite{promislow2014}). If the counterparty has not defaulted
yet, then $\delta\left(t\right)dt$ is the probability that it would
default between $t$ and $t+dt$. The possibility of counterparty
default reduces the value of the option as in case of default the
holder does not receive the whole value of the option. In particular,
the holder can recover only a percentage of the contract value, which
is called the recovery rate $R$. The CVA is defined as the difference
between the risk free price and the risk adjusted price. More precisely,
as stated by Gregory \cite{gregory2010}, the CVA is given by: 
\begin{equation}
\text{CVA}=\left(1-R\right)\int_{0}^{T}D\left(0,s\right)EE\left(s\right)dPD\left(s\right),\label{eq:CVA}
\end{equation}
where $D\left(0,t\right)$ is the risk-free discount factor. This
mean that the CVA is the expected value of the possible losses due
to counterparty default. We stress out that definition requires the
exposure and the counterparties default probability to be independent
and the discount factor to be also independent of the exposure. Using
equations (\ref{eq:EE}) and (\ref{eq:CVA}), we obtain the following
relation: 
\begin{equation}
\text{CVA}=\mathbb{E}\left[\int_{0}^{T}D\left(0,s\right)\left(1-R\right)E\left(s\right)dPD\left(s\right)|\mathcal{F}_{0}\right].\label{eq:CVA_2}
\end{equation}
According to (\ref{eq:CVA_2}), the CVA is the price (mean value of
the future cash flows) of a financial derivative which pays $\left(1-R\right)\max\left[\mathcal{V}\left(\tau_{C},S_{\tau_{C}},V_{\tau_{C}}\right),0\right]$
at time $\tau_{C}$. Therefore, we can consider the CVA as a derivative
itself and we denote its financial value at time $t$ with $\mathcal{C}\left(t,S_{t},V_{t}\right)$,
that is
\[
\mathcal{C}\left(t,S_{t},V_{t}\right)=\mathbb{E}\left[\int_{t}^{T}D\left(0,s\right)\left(1-R\right)E\left(s\right)dPD_{s}|\mathcal{F}_{t}\right],
\]
 having $CVA=\mathcal{C}\left(0,S_{0},V_{0}\right)$. It is possible
to show that $\mathcal{C}\left(t,S_{t},V_{t}\right)$ solves the following
PIDE (see the Appendix for more details):
\begin{equation}
\frac{\partial\mathcal{C}\left(t,S,V\right)}{\partial t}+L_{C}\mathcal{C}\left(t,S,V\right)+L_{J}\mathcal{C}\left(t,S,V\right)+\left(1-R\right)\max\left[\mathcal{V}\left(t,S_{t},V_{t}\right),0\right]\frac{\partial PD}{\partial t}\left(t\right)=0,\label{eq:PIDE_2}
\end{equation}
with the terminal condition 
\begin{equation}
\mathcal{C}\left(T,S,V\right)=0.
\end{equation}

We stress out that equation (\ref{eq:PIDE_2}) depends on the value
$\mathcal{V}\left(t,S_{t},V_{t}\right)$ which has to be computed
previously by solving equation (\ref{eq:PIDE_compact}).

\section{\label{Sec4}Hybrid methods in the Bates model}

The Hybrid Tree Finite Difference (HTFD) and the Hybrid Tree Monte
Carlo (HTMC) methods are two innovative and efficient approaches to
price derivatives when the Bates model is considered. In this Section
we present the main ideas that underlie these two methods: the interested
reader can find more details about the numerical procedures in Briani
et al. \cite{briani2016}.

\subsection{\label{Sec4.1}The HTFD method}

The HTFD method is a backward induction algorithm that works following
a finite difference PIDE method in the direction of the share process
and following a tree method in the direction of the other random sources,
that is volatility in the case of Bates model. Specifically, the method
is based on the following steps. First of all, a binomial tree for
the CIR volatility process $V$ is considered according to Apolloni
et al. \cite{appolloni2014}. Then, a transformation which keeps the
diffusion processes $S$ and $V$ uncorrelated is applied. Finally,
a finite difference approach in the $S$-direction is developed. 

In particular, consider a large integer value $N$, a time horizon
$\left[0,T\right]$ and define $h=T/N$. For $n=0,1,\dots,N$, define
\begin{equation}
V_{n}^{h}=\left\{ v_{n,k}\right\} _{k=0,1,\dots,n}
\end{equation}
 with 
\begin{equation}
v_{n,k}=\left(\sqrt{V_{0}}+\frac{\sigma}{2}\left(2k-n\right)\sqrt{h}\right)^{2}1_{\sqrt{V_{0}}+\frac{\sigma}{2}\left(2k-n\right)\sqrt{h}>0}.
\end{equation}
We define the multiple jumps 
\begin{alignat*}{1}
k_{d}^{h}\left(n,k\right) & =\max\left\{ k^{*}:0\leq k^{*}\leq j\text{ and }v_{n,k}+\mu_{V}\left(v_{n,k}\right)h\geq v_{n+1,k*}\right\} ,\\
k_{u}^{h}\left(n,k\right) & =\min\left\{ k^{*}:k+1\leq k^{*}\leq n+1\text{ and }v_{n,k}+\mu_{V}\left(v_{n,k}\right)h\leq v_{n+1,k^{*}}\right\} 
\end{alignat*}
in which $\mu_{V}$ denotes the drift coefficient of $V$, that is
$\mu_{V}\left(v\right)=\kappa\left(\theta-v\right)$. Starting from
the node $\left(n,k\right)$, the discrete process can reach the up-node
$\left(n+1,k_{u}^{h}\left(n,k\right)\right)$ or the down-jump node
$\left(n+1,k_{d}^{h}\left(n,k\right)\right)$ with transition probability
given by
\begin{align}
\text{up-jump:}\  & p_{\ k_{u}^{h}\left(n,k\right)}^{h}=0\lor\frac{\mu_{V}\left(v_{n,k}\right)h+v_{n,k}-v_{n+1,k_{d}^{h}\left(n,k\right)}}{v_{n+1,k_{u}^{h}\left(n,k\right)}-v_{n+1,k_{d}^{h}\left(n,k\right)}}\wedge1,\label{eq:V_tree}\\
\text{down-jump}:\  & p_{\ k_{d}^{h}\left(n,k\right)}^{h}=1-p_{k_{u}^{h}\left(n,k\right)}^{h}.
\end{align}
Multiple jumps and jump probabilities are set in order to match the
first local moment of the tree and of the process $V$ up to order
one with respect to $h$. As a consequence, as $h$ approaches to
0, the weak convergence on the path space is guaranteed. Moreover,
in order to obtain the convergence, the Feller condition (Albrecher
et al. \cite{albrecherj2007}) is not required.

Let us consider the diffusion pair $\left(Y,V\right)$, where $Y$
is a stochastic process defined by
\begin{equation}
Y_{t}=\log\left(S_{t}\right)-\frac{\rho}{\sigma}V_{t}.\label{eq:trasformazione_S_Y}
\end{equation}
Clearly the couple $\left(S,V\right)$ can be retraced by $\left(Y,V\right)$
by inverting relation (\ref{eq:trasformazione_S_Y}). We set $\bar{\rho}=\sqrt{1-\rho^{2}}$
and we consider $\left(W,Z\right)$ as a standard Brownian motion
in $\mathbb{R}^{2}$. Then, the dynamics of the couple $\left(Y,V\right)$
is given by 
\begin{align}
dY_{t} & =\left(r-\eta-\frac{1}{2}V_{t}-\frac{\rho}{\sigma}\kappa\left(\theta-V_{t}\right)\right)dt+\bar{\rho}\sqrt{V_{t}}dZ_{t}+dN_{t},\\
dV_{t} & =\kappa\left(\theta-V_{t}\right)dt+\sigma\sqrt{V_{t}}dW_{t},
\end{align}
 with $Y_{0}=\log S_{0}-\frac{\rho}{\sigma}V_{0}$. Here, $N_{t}$
is the compound Poisson process written through the Poisson process
$K$, with intensity  $\lambda$, and the i.i.d. jumps $\left\{ \log\left(1+J_{k}\right)\right\} $,
that is 
\begin{equation}
N_{t}=\sum_{k=1}^{K_{t}}\log\left(1+J_{k}\right).
\end{equation}
We set 
\begin{equation}
\mu_{Y}\left(v\right)=r-\eta-\frac{1}{2}v-\frac{\rho}{\sigma}\kappa\left(\theta-v\right)
\end{equation}
 and 
\begin{equation}
\mu_{V}\left(v\right)=\kappa\left(\theta-v\right).
\end{equation}
Let $\bar{V}^{h}=\left(\bar{V}_{n}^{h}\right)_{n=0,\dots,N}$ denote
the tree process approximating $V$ and set $V_{t}^{h}=\bar{V}_{\left\lfloor t/h\right\rfloor }^{h},$
$t\in\left[0,T\right]$, the associated piecewise constant and càdlàg
approximation path. In order to approximate $Y$, Briani et al. construct
a Markov chain from the finite difference method. Starting from the
Euler scheme $Y_{0}^{h}=Y_{0}$ and for $t\in\left(nh,\left(n+1\right)h\right]$,
$n=0,\dots,N$, it is possible to set 
\begin{equation}
Y_{t}^{h}=Y_{nh}^{h}+\mu_{Y}\left(V_{nh}^{h}\right)\left(t-nh\right)+\bar{\rho}\sqrt{V_{nh}^{h}}\left(Z_{t}-Z_{nh}\right)+\left(N_{t}-N_{nh}\right),\label{eq:Y_tree}
\end{equation}
 $Z$ being independent of the noise driving $\bar{V}^{h}$. Now,
let $\mathcal{W}\left(t,Y,V\right)=\mathcal{V}\left(t,\exp\left(Y+\frac{\rho}{\sigma}V\right),V\right)$
the function which gives the price of the financial derivative at
time $t$ in terms of the couple $\left(Y,V\right)$ . Then,
\begin{align}
\mathbb{E}\left(\mathcal{W}\left(\left(n+1\right)h,Y_{\left(n+1\right)h},V_{\left(n+1\right)h}\right)|Y_{nh}=y,V_{nh}=v\right) & \approx\mathbb{E}\left(\mathcal{W}\left(\left(n+1\right)h,Y_{\left(n+1\right)h}^{h},V_{\left(n+1\right)h}^{h}\right)|Y_{nh}=y,V_{nh}=v\right)\\
 & =\mathbb{E}\left(u^{h}\left(nh,y;v,V_{\left(n+1\right)h}^{h}\right)|V_{nh}^{h}=v\right)
\end{align}
where 
\begin{equation}
u^{h}\left(nh,y;v,z\right)=\mathbb{E}\left(\mathcal{W}\left(\left(n+1\right)h,Y_{\left(n+1\right)h}^{h},z\right)|Y_{nh}^{h}=y,V_{nh}^{h}=v\right)
\end{equation}
 and 
\begin{equation}
u^{h}\left(nh,y;v,z\right)=u^{h}\left(s,y;v,z\right)|_{s=nh}.
\end{equation}
The key point of the HTFD method is that the function $\left(s,y\right)\mapsto u^{h}\left(s,y;v,z\right)$
solves the following PIDE in the time interval $nh<s<\left(n+1\right)h$:
\begin{equation}
\frac{\partial u^{h}}{\partial s}+\mu_{Y}\left(v\right)\frac{\partial u^{h}}{\partial y}+\frac{1}{2}\bar{\rho}^{2}v\frac{\partial^{2}u^{h}}{\partial y^{2}}+\int_{-\infty}^{+\infty}\left[u^{h}\left(s,y+x;v,z\right)-u^{h}\left(s,y;v,z\right)\right]p_{J}\left(x\right)dx=0,\label{eq:PIDE_u}
\end{equation}
 for $y\in\mathbb{R}$, and with terminal condition given by
\begin{equation}
u^{h}\left(\left(n+1\right)h,y;v,z\right)=\mathcal{W}\left(\left(n+1\right)h,y,v\right).\label{eq:terminal_u}
\end{equation}

In order to numerically solve the PIDE (\ref{eq:PIDE_u}) by using
a finite difference scheme, we first localize the variables and the
integral term to bound the computational domains. For this purpose,
we use the estimates for the localization domain and the truncation
of large jumps given by Voltchkova and Tankov \cite{voltchkova2008}.
Then, the derivatives of the solution are replaced by finite differences
and the integral terms are approximated using the trapezoidal rule.
Finally, the problem is solved by using an explicit-implicit scheme.

We observe that the above algorithm is referred to a European option,
can easily be adapted to consider an American option. Specifically,
we approximate the American option with a Bermudan option with exercise
dates given by the set $\left\{ nh\right\} _{n=0,\dots,N}$ and replacing
equation (\ref{eq:terminal_u}) with the following one: 
\begin{equation}
u^{h}\left(\left(n+1\right)h,y;v,z\right)=\max\left[\mathcal{W}\left(\left(n+1\right)h,y,v\right),\psi\left(\exp\left(y+\frac{\rho}{\sigma}v\right)\right)\right]\label{eq:terminal_u-1}
\end{equation}

We stress out that the computation of $u^{h}$ as the solution of
(\ref{eq:PIDE_u}) is a one dimensional problem with constant coefficients,
thus it can be solved in a very efficient way, with a low computational
cost.

\subsection{The HTMC method}

The HTMC method is a Monte Carlo algorithm which is based on the approximations
(\ref{eq:V_tree}) and (\ref{eq:Y_tree}). In particular, the HTMC
method consists in simulating a continuous process in space (the component
$Y$) starting from a discrete process in space (the 1-dimensional
tree for $V$). In particular, we set $\hat{Y}_{0}^{h}=Y_{0}$ for
$t\in\left[nh,\left(n+1\right)h\right]$ with $n=0,1,\dots,N-1$.
Then, we compute $\hat{Y}_{n+1}^{h}$ recursively by the following
relation:
\begin{equation}
\hat{Y}_{n+1}^{h}=\hat{Y}_{n}^{h}+\mu_{Y}\left(\hat{V}_{n}^{h}\right)h+\hat{\rho}\sqrt{h\hat{V}_{n}^{h}}\Delta_{n+1}+\left(N_{\left(n+1\right)h}-N_{nh}\right)\label{eq:ricorsiva}
\end{equation}
where $\Delta_{1},\dots,\Delta_{N}$ are i.i.d. standard normal random
variables which are independent of the noise driving the Markov chain
$\hat{V}$ and $\left(N_{\left(n+1\right)h}-N_{nh}\right)$ is the
compound Poisson increment. Roughly speaking, one let the pair $\left(Y,V\right)$
evolve on the tree and simulate the process Y at time $nh$ by using
relation (\ref{eq:ricorsiva}).

\section{Numerical method for computing the CVA}

In this section we present the proposed approach to compute the CVA,
namely the Coupled-Hybrid Tree Finite Differences (C-HTFD), which
is based on the resolution of two coupled PIDE. Moreover, we start
presenting the Hybrid Tree Finite Difference-Hybrid Monte Carlo (HTFD-HTMC)
method which is based on the direct application of the hybrid techniques
in the original FDMC (we consider this method for comparison purposes
mainly). The two approaches are both based on the application of the
hybrid algorithms of Briani et al. \cite{briani2016}, presented in
Section \ref{Sec4}.

\subsection{The HTFD-HTMC approach }

This method is based on the direct application of the hybrid techniques
to the Finite Difference Monte Carlo (FDMC) method, which was first
developed by de Graaf et al. \cite{deGraaf2017} for the Heston model
and then adapted for the Bates model by Feng \cite{feng2017}. First
of all, an estimation of the value function $\mathcal{V}\left(t,S,V\right)$
is computed through a grid of values by solving equation (\ref{eq:PIDE})
by employing the Alternating Direction Implicit (ADI) method and specifically
the scheme proposed by Haentjens and In't Hout \cite{haentjens2012}.
Then a Monte Carlo simulation is employed to estimate the expectation
in (\ref{eq:CVA}) and thus estimate the CVA.

In the HTFD-HTMC approach we employ the HTFD method to solve (\ref{eq:PIDE})
and the HTMC method to estimate (\ref{eq:CVA}). Specifically, let
$h=T/N$ as in Subsection \ref{Sec4.1}. First of all, the HTFD method
is used to compute the risk free price $\left\{ \mathcal{V}\left(nh,S_{nh},V_{nh}\right)\right\} _{n=0,\dots,N}$
at discrete times $\left\{ 0,h,\dots,Nh\right\} $. Then, we estimate
the expected exposure (\ref{eq:EE}) via a set of $N_{MC}$ Monte
Carlo simulations $\left\{ \left(S_{nh}^{j},V_{nh}^{j}\right),n=0,\dots,N\text{ and }j=1,\dots,N_{MC}\right\} $
generated via the HTMC method, that is 
\begin{equation}
EE\left(nh\right)\approx\frac{1}{N_{MC}}\sum_{j=1}^{N_{MC}}\max\left[\mathcal{V}\left(nh,S_{nh}^{j},V_{nh}^{j}\right)\right].
\end{equation}
 Finally, we compute the CVA by approximating the integral in (\ref{eq:CVA})
using the uniform partition $\left\{ 0,h,\dots,Nh\right\} $ of $\left[0,T\right]$,
the estimated value $\left\{ EE\left(nh\right),n=0,\dots,N\right\} $
and the trapezoidal rule.

\subsection{The C-HTFD approach }

First of all, an estimation of the value function $\mathcal{V}\left(t,S,V\right)$
is computed through a grid of values by solving equation (\ref{eq:PIDE})
by means of the HTFD method. Then, the CVA is computed as the solution
of the PIDE (\ref{eq:PIDE_2}), which is done again by employing the
HTFD method. We stress that C-HTFD method and the HTFD-HTMC method
share the first step, but in the C-HTFD method the Monte Carlo step
is replaced with a second finite difference like step.

\section{Numerical Results}

In this section we propose the results of some numerical experiments,
which aim to compare the goodness of the proposed methods. In particular,
we compare the standard  FDMC method employed by Feng \cite{feng2017},
the HTFD-HTMC approach and the C-HTFD approach. We employ the numerical
methods according to 4 configurations (\emph{A, B, C, D}), each of
them with an increasing number of steps, determined in order to achieve
approximately these run times\footnote{We performed the numerical tests using a personal computer with the
following features. CPU: Intel(R) Core(TM) i5-7200 2.50 GHz; RAM:
8GB, DDR4} : $\left(A\right)$ 0.25 s, $\left(B\right)$ 1 s, $\left(C\right)$
4 s, $\left(D\right)$ 16 s. The employed mesh and configuration parameters
are reported in Table \ref{tab:Configuration-parameters}. Specifically,
for the FDMC method with the sequence (time steps - points for $S$
- points for $V$ - MC simulations) and for the hybrid methods with
the sequence (time steps - points for $Y$ - MC simulations). Moreover,
we reported the $95\%$ confidence intervals for the FDMC method and
for the HTFD-HTMC. Furthermore, we employ the FDMC method with a huge
number of steps and paths $\left(250;500;200;10^{6}\right)$ as a
benchmark (\emph{BM}).

We consider the following parameters for the Bates model: $S_{0}=80,\ 100,\ 120$,
$K=100$ , $T=1$, $r=0.03$, $\eta=0.00$, $V_{0}=0.01$ $\kappa=2$,
$\theta=0.01$, $\sigma=0.2$, $\lambda=0.1$, $\alpha=0.1$, $\beta^{2}=0.1$
and $\rho=0.5$. According to the default parameters, we consider
$\delta\left(t\right)=\delta=0.03$ and $R=0.4$. We compute the CVA
for a European and an American a Put option with strike equal to $K=100$.

Results are available in Tables \ref{tab:results} and \ref{tab:results-1},
which show that all the implemented methods give similar results.
Values calculated via HTFD-HTMC and values calculated via FDMC have
similar accuracy. We emphasize that, despite the similarity in numerical
results of these two methods, we prefer the HTFD-HTMC method because
of its simplicity of implementation.

The values returned by C-HTFD are the most accurate since they are
very close to the benchmark for the whole parameter configurations. 

\begin{table}[p]
\begin{centering}
\begin{tabular}{cccc}
\toprule 
 & FDMC & HTFD-HTMC & C-HTFD\tabularnewline
\midrule
A & $\phantom{1}50;80;15;1500$ & $\phantom{1}50;100;1500$ & $\phantom{1}50;100$\tabularnewline
B & $\phantom{1}75;110;30;2000$ & $\phantom{1}75;150;2000$ & $\phantom{1}75;150$\tabularnewline
C & $100;200;50;3300$ & $100;250;3300$ & $100;250$\tabularnewline
D & $125;300;100;6000$ & $125;350;6000$ & $125;350$\tabularnewline
\bottomrule
\end{tabular}
\par\end{centering}

\caption{\label{tab:Configuration-parameters} Configuration parameters for
the numerical methods.}
\end{table}

\begin{table}[p]
\begin{centering}
\begin{tabular}{cccccc}
\toprule 
 &  & FDMC & HTFD-HTMC & C-HTFD & BM\tabularnewline
\midrule
\multirow{4}{*}{$S_{0}=80$} & A & $0.320071\pm0.005172$ & $0.327462\pm0.006098$ & $0.323732$ & \multirow{4}{*}{$0.323724\pm0.000200$}\tabularnewline
 & B & $0.322874\pm0.004469$ & $0.327165\pm0.005107$ & $0.323713$ & \tabularnewline
 & C & $0.323805\pm0.003435$ & $0.326023\pm0.003993$ & $0.323707$ & \tabularnewline
 & D & $0.324926\pm0.002660$ & $0.324871\pm0.002911$ & $0.323703$ & \tabularnewline
\midrule
\multirow{4}{*}{$S_{0}=100$} & A & $0.058209\pm0.003071$ & $0.063808\pm0.003809$ & $0.060724$ & \multirow{4}{*}{$0.060359\pm0.000125$}\tabularnewline
 & B & $0.059096\pm0.002635$ & $0.062193\pm0.003005$ & $0.060613$ & \tabularnewline
 & C & $0.059838\pm0.002178$ & $0.061673\pm0.002385$ & $0.060507$ & \tabularnewline
 & D & $0.060978\pm0.001664$ & $0.060749\pm0.001678$ & $0.060467$ & \tabularnewline
\midrule
\multirow{4}{*}{$S_{0}=120$} & A & $0.005658\pm0.001697$ & $0.007251\pm0.001976$ & $0.005633$ & \multirow{4}{*}{$0.005589\pm0.000059$}\tabularnewline
 & B & $0.005406\pm0.001289$ & $0.006296\pm0.001461$ & $0.005610$ & \tabularnewline
 & C & $0.005334\pm0.000954$ & $0.006299\pm0.001328$ & $0.005592$ & \tabularnewline
 & D & $0.005660\pm0.000758$ & $0.005616\pm0.000852$ & $0.005584$ & \tabularnewline
\bottomrule
\end{tabular}
\par\end{centering}
\caption{\label{tab:results}CVA for European Put options.}
\end{table}

\begin{table}[p]
\begin{centering}
\begin{tabular}{cccccc}
\toprule 
 &  & FDMC & HTFD-HTMC & C-HTFD & BM\tabularnewline
\midrule
\multirow{4}{*}{$S_{0}=80$} & A & $0.336835\pm0.005078$ & $0.342798\pm0.006316$ & $0.338987$ & \multirow{4}{*}{$0.339054\pm0.000208$}\tabularnewline
 & B & $0.337859\pm0.004630$ & $0.342703\pm0.005295$ & $0.339084$ & \tabularnewline
 & C & $0.338970\pm0.003560$ & $0.341533\pm0.004139$ & $0.339134$ & \tabularnewline
 & D & $0.340182\pm0.002755$ & $0.340498\pm0.003019$ & $0.339165$ & \tabularnewline
\midrule
\multirow{4}{*}{$S_{0}=100$} & A & $0.059803\pm0.003168$ & $0.065669\pm0.003937$ & $0.062466$ & \multirow{4}{*}{$0.062145\pm0.000130$}\tabularnewline
 & B & $0.060742\pm0.002723$ & $0.063979\pm0.003103$ & $0.062364$ & \tabularnewline
 & C & $0.061522\pm0.002248$ & $0.063446\pm0.002458$ & $0.062260$ & \tabularnewline
 & D & $0.062717\pm0.001718$ & $0.062501\pm0.001732$ & $0.062221$ & \tabularnewline
\midrule
\multirow{4}{*}{$S_{0}=120$} & A & $0.005798\pm0.001745$ & $0.005449\pm0.002043$ & $0.005782$ & \multirow{4}{*}{$0.005740\pm0.000061$}\tabularnewline
 & B & $0.005537\pm0.001327$ & $0.006466\pm0.001509$ & $0.005760$ & \tabularnewline
 & C & $0.005473\pm0.000983$ & $0.006467\pm0.001368$ & $0.005742$ & \tabularnewline
 & D & $0.005812\pm0.000784$ & $0.005763\pm0.000877$ & $0.005735$ & \tabularnewline
\bottomrule
\end{tabular}
\par\end{centering}
\caption{\label{tab:results-1}CVA for American Put options.}
\end{table}

\section{Conclusions}

In this paper we have proposed a numerical method to compute the CVA
of European and American options when the underlying is assumed to
evolve according to the Bates model. This method is based on the the
resolution of two coupled PIDE which is done by employing the Hybrid
Tree-Finite Difference algorithm developed by Briani et al.. Specifically,
the C-HTFD approach consist in replacing the Monte Carlo step of the
FDMC method with the resolution of the PIDE followed by the CVA cost,
which is done by employing the Hybrid technique for the Bates model.

Numerical results show that our method is very stable and robust.
In particular, numerical tests reveals that the values returned by
C-HTFD method are very accurate, much more than the results provided
by other methods which involve a Monte Carlo step, since the use of
a PIDE approach in place of a Monte Carlo one dramatically improves
the computational efficiency. Thus, the C-HTFD is efficient and reliable
and the use of the two coupled PIDE represents a relevant improvement
with respect to the standard pricing techniques of the CVA when the
Bates model is considered.

\newpage \clearpage

\begin{appendices} 

\section{Proof of PIDE (\ref{eq:PIDE_2})}

This appendix provides a derivation of the PIDE (\ref{eq:PIDE_2})
followed by the the CVA price. First of all, we present the proof
for the simple case of an underlying evolving according to the Black-Scholes
model. Then we consider the Bates model.

\subsection*{Black-Scholes model}

Let $S$ denote the underlying, which is assumed to evolve according
to 
\begin{equation}
\frac{dS_{t}}{S_{t}}=(r-\eta)dt+\sigma\,dW_{t},
\end{equation}
 where $\eta$ denotes the continuous dividend rate, $S_{0}$ is a
positive value and $W_{t}$ is a Brownian motion. We assume that counterparty
credit risk is diversifiable across a large number of counterparties.
In the case that this assumption is not justified, then the risk-neutral
value of the contract can be adjusted using an actuarial premium principle
(Gaillardetz and Lakhmiri \cite{gaillardetz2011}). Moreover, the
fraction of the original counterparties of the contract who have defaulted
before time $t$ is given by 
\begin{equation}
PD\left(t\right)=1-\exp\left(-\int_{0}^{t}\delta\left(s\right)ds\right).
\end{equation}

Let us consider a financial product which pays $\left(1-R\right)E\left(t\right)$
if the counterparty defaults at time $t$ and $0$ otherwise. Then
the value of such a product at time $0$ - the discounted value of
future cash-flows - is equal to the CVA. Therefore, we can consider
the CVA as a derivative itself and we denote its financial value at
time $t$ with $\mathcal{C}\left(t,S_{t},V_{t}\right)$, that is
\begin{equation}
\mathcal{C}\left(t,S_{t},V_{t}\right)=\mathbb{E}\left[\int_{t}^{T}D\left(0,s\right)\left(1-R\right)E\left(s\right)dPD_{s}|\mathcal{F}_{t}\right],
\end{equation}
 having $CVA=\mathcal{C}\left(0,S_{0},V_{0}\right)$. 

Suppose that the writer of a CVA derivative forms a self-financing
portfolio portfolio $\Pi$ which, in addition to being short to the
CVA, is long $x$ units of the index $S$, i.e.
\begin{equation}
\Pi=-\mathcal{C}\left(t,S,V\right)+xS.
\end{equation}

Then, by It\^o's lemma,
\begin{equation}
d\Pi=-\left[\frac{\partial\mathcal{C}}{\partial t}+\frac{\sigma^{2}S^{2}}{2}\frac{\partial^{2}\mathcal{C}}{\partial S^{2}}+\mu S\frac{\partial\mathcal{C}}{\partial S}\right]dt-\sigma S\frac{\partial\mathcal{C}}{\partial S}dW+x\mu Sdt+x\sigma SdW-\frac{dPD\left(t\right)}{dt}\left(1-R\right)E\left(t\right)dt\label{eq:dPi}
\end{equation}
where the last term reflects the cash-flows from the fraction of the
original counterparties who default between $t$ and $t+dt$. Setting
$x=\frac{\partial\mathcal{C}}{\partial t}$ in equation (\ref{eq:dPi})
gives 
\begin{equation}
d\Pi=-\left[\frac{\partial\mathcal{C}}{\partial t}+\frac{\sigma^{2}S^{2}}{2}\frac{\partial^{2}\mathcal{C}}{\partial S^{2}}\right]dt-\frac{dPD\left(t\right)}{dt}\left(1-R\right)E\left(t\right)dt.
\end{equation}
Since the portfolio is now (locally) riskless, it must earn the risk-free
interest rate $r$. Setting $d\Pi=r\Pi dt$ results in 
\begin{equation}
-\left[\frac{\partial\mathcal{C}}{\partial t}+\frac{\sigma^{2}S^{2}}{2}\frac{\partial^{2}\mathcal{C}}{\partial S^{2}}\right]dt-\frac{dPD\left(t\right)}{dt}\left(1-R\right)E\left(t\right)dt=r\left[-\mathcal{C}+\frac{\partial\mathcal{C}}{\partial t}S\right]dt.
\end{equation}
Therefore, we get
\begin{equation}
\frac{\partial\mathcal{C}}{\partial t}+\frac{\sigma^{2}S^{2}}{2}\frac{\partial^{2}\mathcal{C}}{\partial S^{2}}+rS\frac{\partial\mathcal{C}}{\partial S}-r\mathcal{C}+\left(1-R\right)E\left(t\right)\frac{dPD\left(t\right)}{dt}=0.
\end{equation}

\subsection*{Bates model}

The proof of PIDE (\ref{eq:PIDE_2}) in the case of the Bates model
may be done by employing the same approach followed in the previous
Sub Section for the Black-Scholes model. Alternatively, we can obtain
the PIDE (\ref{eq:PIDE_2}) by a straightforward application of the
Feynman\--Kac formula for L\'evy processes (Rong \cite{rong1997},
Glau \cite{glau2016}), which can be applied since the CVA is defined
as the expected value of a time integral of function depending on
a L\'evy process.

\end{appendices}

\bibliographystyle{abbrv}
\bibliography{bibliography}

\end{document}